# Hydrothermal synthesis, structure, and antibacterial studies of a nanosized iron zeolite

Karam S. El-Nasser[a,b*], T.A. Taha[c,d], Ibraheem O. Ali[a,e], Hossam Donya[f,g]

[a]*Chemistry Department, College of Science and Arts, jouf University, P.O. Box 756, Al-Gurayyat, Saudi Arabia* [b]*Chemistry Department, Faculty of Science, Al-Azhar University, Assiut, 71524, Egypt*

[c]*Physics Department, College of Science and Arts, Jouf University, P.O. Box 756, Al-Gurayyat, Saudi Arabia*

[d]*Physics and Engineering Mathematics Department, Faculty of Electronic Engineering, Menoufia University, Menouf, 32952, Egypt*

[e] Department of Chemistry, Faculty of Science, Al-Azhar University, Nasr City 11884, Cairo, Egypt

[f]*Physics Department, Faculty of Science, Menoufia University, Shebin El-Koom, Egypt* [g]*Department of Physics, Faculty of science, King Abdulaziz University, Jeddah 21589, Saudi Arabia*

**\*Corresponding author. College of Sciences and Arts, Jouf University, P.O. Box 756, Al-Gurayat, Saudi Arabia. Tel.: +966-533472910. Email address; karamsaif@ju.edu.sa**

## Abstract

The present research explores the effects of %Fe incorporation on the structure and antibacterial activity of Fe-ZSM-5. Silica extracted from local rice husk straw (white particles) by applying NaOH and HCl solutions for consecutive chemical treatment was used for hydrothermal synthesis of FeZSM-5 with constant Si/(Fe + Al) ratios. The chemical and physical changes of ZSM-5 and Fe/ZSM-5 surfaces were investigated by X-ray diffraction analysis (XRD), scanning electron microscopy (SEM), Fourier transform infrared spectroscopy (FTIR), UV-Vis spectroscopy, differential scanning calorimetry (DSC), and pore structure analysis by $N_2$ adsorption at -196 °C. XRD analysis revealed the typical ZSM-5 structure with new diffraction lines attributed to the iron silicate phase. FTIR spectral analysis of ZSM-5 samples containing iron display a new band at 656 cm$^{-1}$ that is ascribed to the Si–O– Fe group. The antibacterial activity of such coatings towards different kinds of bacteria, such as *S. pneumonia*, *B. subtilis*, *E. coli* and *P. aeruginosa*, and fungi, such as *A. fumigatus* and *C. albicans*, for investigated ZSM-5 and Fe (20% and 100%) samples showed selective antibacterial actions.

# 1. Introduction

Zeolites are widely utilized in three fundamental applications: adsorbents, catalysts, and ion exchange materials. The benefits of zeolites that that make them suitable for these applications include the heterogeneous nature of the catalysts for easy separation, the ability to be doped with metals to achieve selective oxidation chemistry and ease of catalyst regeneration [1-4]. The possibility of organic transformations and separations coupled in one unit is offered by zeolitic membranes. Redox molecular sieves were expected to be used in fine chemical synthesis, exploiting the considerable flexibility in both designing the topology framework and inserting reactive elements and compounds into the framework. Other specialized applications include fabrication of sensors, photochemical organic transformations, and conversion of solar energy [5-7].

Extensive work on ZSM-5 synthesis and applications has been carried out since the discovery of this material. In an autoclave at 120–180 ºC, highly crystalline ZSM-5 was synthesized under autogenous pressures within a period of approximately 8 days. The reaction time was decreased to 4–6 h under high temperatures and pressures (230–250 ºC and 40–60 atm) [8-10]. Rice husk (RH) is a solid waste that contains 41% (w/w) amorphous silica and is suitable for recycling from undesirable agricultural mass residues. Unwashed husk ash, however, contains approximately 96% (w/w) silica and some organics, alkali oxides, and impurities. Additionally, amorphous silica extracted from inexpensive rice straw ash could be used to produce siliconbased materials for industrial and technological applications [11,12]. Elements with an appropriate size (ionic radius) for stable tetrahedral coordination, such as, Al and Ge and, to a lesser extent, Be, Ga, Fe and B, are suitable for extended incorporation into the zeolite structure. Other more voluminous elements, such as Zr, V, Cr, Mo,

Sn, In, Zn and Ti, are less readily accommodated [13–17]. Incorporation of Fe into the zeolite structure has attracted increasing attention recently for wide-ranging catalytic applications [18,19], e.g., methanol oxidation [20], oxidation of propane [21] and decomposition of n-butane [22] and $N_2O$ without any reduction agent [23]. The development of materials that are able to inhibit microbial growth, such as water purification systems, medical devices, food packaging and storage systems, construction and textiles [24-27], has attracted great attention in academic and technological fields. Diverse work has shown that the presence of inorganic antimicrobial additives has significant benefits over conventional organic agents due to the high chemical stability, thermal resistance, improved user safety, and long-lasting duration of action associated with these materials [28-30].

Iron-ZSM-5 was prepared from rise husk straw (RHS) in this study through replacement of Al by Fe. The prepared samples were investigated via XRD, FTIR, DSC, UV-Vis, and $N_2$ adsorption. Additionally, these samples exhibited antimicrobial activity.

## 2. Experimental

### 2.1. Materials

We used RHS, NaOH pellets (AR, 98%), $Al_2(SO_4)_3.16H_2O$ (Merck), tetrapropyl ammonium bromide (TPAB) (Fluka), n-propyl amine (*n*-PA) (Merck), $H_2SO_4$, and $Fe(NO_3)_3 \cdot 9H_2O$ (Merck).

### 2.2. Silica extraction from RHS

RHS was washed, air dried at 300 K, and heated for 1 h under reflux in a suitable amount of hydrochloric acid (3%). The filtered slurry was washed with distilled water to remove chloride ions using a $AgNO_3$ solution, dried at 110 °C, and calcinated for 6 h at 750 °C to obtain RHS-HCl white ash. As determined from X-ray fluorescence (XRF: Philips PW1400) measurements, the silica yield in the obtained sample was 42%, and the chemical composition analysis yielded the following (on a wt % basis): 4.71; $SiO_2$, 90.70; $Al_2O_3$, 0.13; $Fe_2O_3$, 0.06; $TiO_2$, 0.015; CaO, 0.61; MgO, 0.25; $Na_2O$, 0.09; $K_2O$, 2.64; $P_2O_5$, 0.73.

## 2.3. Synthesis of ZSM-5 zeolite

In the hydrothermal preparation of ZSM-5 zeolite, the hydrogel was prepared as follows: $3.21Na_2O:0.162Al_2O_3:6.185SiO_2:0.185Q:100H_2O$, where Q refers to the TPABr ($C_{12}H_{28}NBr$) template, as mentioned in reference [1].

## 2.4. Synthesis of Fe-ZSM-5

The samples have the following composition (molar ratio):

$3:12Na_2O:(1-x)Al_2O_3:xFe_2O_3:6.185SiO_2:0.185TPABr:100H_2O$, where $x$ varied from 0.0 to 0.8 in each series (100% Fe).

Certain quantities of iron nitrate solution were added dropwise to a mixture of TPABr, n-propyl amine, sodium silicate, and aluminium sulfate (pH = 11) with constant stirring for 30 min. The obtained gel was aged for 1 h and then transferred to an autoclave under autogenous pressure at 160 °C in an oil bath. After 8 days, the autoclave was removed from the oil bath and quenched in cold water for product identification. The products were filtered, washed, and dried at 120 °C for 10 h, and the template was removed by sample calcination at 550 °C for 6 h. All sample series were synthesized from gels with a total $SiO_2/(Al_2O_3 + Fe_2O_3)$ molar ratio of 38 and were referred to as 0.0%Fe, 20%Fe, 40%Fe, 60%Fe, 80%Fe, and 100%Fe depending on the Fe ratio.

## 2.5. Physico-chemical characterizations

Sample crystallinity was analysed by XRD on a Philips 3710 diffractometer (CuKα = 1.5404 Å) at a scanning speed of 2θ = 2.5° $min^{-1}$. Surface morphology of the present samples has been assessed by measuring SEM micrographs using FE-SEM electron microscope, Quanta FEG 250, Holland. Characteristic FTIR spectra were measured via a Bruker Vector 22 instrument at a resolution of 2 $cm^{-1}$. We measured the UV-Vis spectra on a JASCO V570 instrument with a bandwidth of 2 nm. DSC thermograms were obtained for the current samples using a Shimadzu DTA-50 instrument at up to 1000 °C.

A volumetric apparatus was used to obtain nitrogen adsorption isotherms of the present samples at -196 °C (0.1 g of the sample was outgassed under $10^{-5}$ Torr and at 300 °C for 3 h). The Brunauer–Emmett–Teller (BET) method was applied to estimate the specific surface area ($S_{BET}$), while the "$t$-plot" method was applied to estimate the micropore volume ($V_P^\mu$) and the external surface area ($S^{ext}$) [31].

## 2.6. Antimicrobial screening

Two gram-negative bacteria—*Escherichia coli* NCTC 10416 and *Pseudomonas aeruginosa* NCIB 9016—and two gram-positive bacteria—*Bacillus subtilis* NCIB 3610 and *Staphylococcus aureus* NCTC 7447—were used to test the antibacterial activity of the present samples, and *Candida albicans* IMRU 3669 was used to evaluate the anti-fungal activity using nutrient agar. Sterilized medium (40-50 °C) was incubated (1 ml/100 ml of medium) with the microorganism suspension (105 cfu ml$^{-1}$) (matched to the McFarland barium sulfate standard) and poured into a glass dish to a depth of 3-4 mm. A paper placed on the solidified medium was impregnated with the test compounds (μg/ml$^{-1}$ in methanol). At room temperature, the plates were pre-incubated for 1 h and incubated to evaluate the antibacterial and antifungal activities at 37 °C for 24 and 48 h, respectively. Antibacterial and antifungal ampicillin (mg/disc) was used as a standard.

Results and discussion

The XRD patterns displayed in Fig. 1 revealed the typical ZSM-5 structure with new diffraction lines attributed to the iron silicate phase, which appeared weakly at d=5.3497, 4.696, 4.469, 4.0736, 3.996, 3.7120, 3.39, 2.9426 and 1.5996 Å, with a small fraction of $Fe_2O_3$ at d=2.514 Å. The size of the $Si_3O_6$ crystals estimated from the Scherrer equation (D= $0.9\lambda/\beta\cos\theta$) [32-37] is 47 nm, which is sufficiently large to fit within the pores formed by the ZSM-5 channel intersections located on the external surface. Of particular interest, this shift to higher d values for ZSM-5 for all levels of Fe substitution means that Fe is well incorporated. Moreover, when Al was gradually replaced by Fe, the peak intensity ratio at $2\theta$ =8-10 and 20-25º was also enhanced. It is well known that these ratios depend sensitively on the $SiO_2/Al_2O_3$ ratio for aluminosilicate zeolite. It is therefore suggested that changes in the intensity ratios between different Fe-containing ZSM-5 are mainly due to increases in the ratio of
$SiO_2/Al_2O_3$ [38].

Table 1 summarizes the Fe-ZSM-5 sample lattice parameters (a, b, c) and unit cell volume (V). The corresponding parameters of the Al-ZSM-5 samples with Al replaced by Fe were improved compared to those of the 60% Fe and 80% Fe samples, suggesting the favourability of the successful combination of Fe in the ZSM-5 network at lower concentrations (20% and 40%). This result occurs because the Si-O and Fe-O bond distances are 0.160 and 0.179 nm, respectively, as in pure Fe-silicate zeolite (100% Fe) [38].

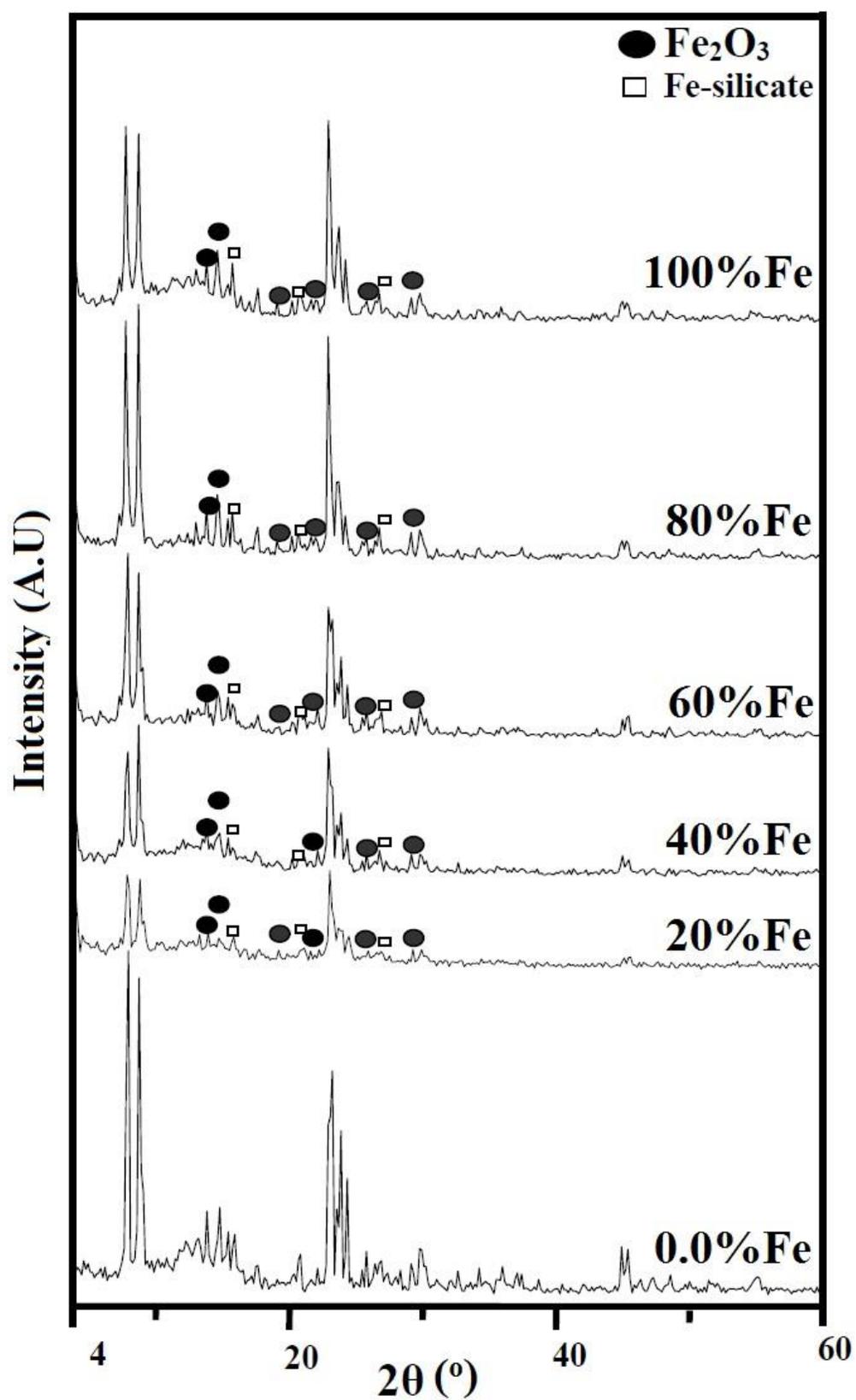

**Fig. 1** XRD spectra of the Fe-ZSM-5 samples

A peak was observed at 2θ = 22-25º for the ZSM-5 crystallites in all samples, and the average particle size values estimated via the Scherrer equation were smaller than those of the parent sample. This result showed that the incorporation of iron prevents particle agglomeration, allowing the materials to maintain their dispersion and thus confirming the expected strong interactions between Fe and ZSM-5.

**Table 1** Values of particle size and lattice parameters for the Fe-ZSM-5 samples

| Sample | D (Å) | Cell parameters | | | V (Å)$^3$ | Crystal. by XRD % | Crystal. by FTIR % |
|---|---|---|---|---|---|---|---|
| | | a | b | c | | | |
| 0.0%Fe | 99.28 | 19.652 | 19.818 | 13.363 | 5204.440 | 100 | 100 |
| 20%Fe | 72.18 | 20.259 | 20.010 | 13.380 | 5424.019 | 69 | 70 |
| 40%Fe | 86.16 | 19.920 | 19.906 | 13.345 | 5291.659 | 76 | 74 |
| 60%Fe | 60.24 | 19.735 | 19.558 | 13.370 | 5160.252 | 60 | 75 |
| 80%Fe | 75.96 | 19.812 | 19.501 | 13.315 | 5144.037 | 83 | 85 |
| 100%Fe | 69.22 | 20.864 | 19.833 | 13.393 | 5541.965 | 59 | 44 |

On the other hand, the ZSM-5 samples with 60% and 80% replacement of Al with Fe have the lowest volume, which decreases the diffusion of Fe ions within channels in the ZSM-5 structure. Subsequently, intense peaks attributed to $Fe_2O_3$ species appeared; in comparison, the 20% and 40% FeZSM-5 samples exhibited a crystallinity comparable to that of the 20%FeZSM-5 sample (30%). Because the strongest peaks at $2\theta = 22–25°$ were increased in the 100%Fe sample, Fe ion diffusion within zeolite channels did not occur; however, the crystallinity of other samples containing 20% and 40% Fe decreased. Thus, the former sample's increase in crystallinity could be due to iron silicate [$Fe_2(Si_2O_7)$] phase formation [13].

As shown in Figure 1, SEM micrographs of Fe-ZSM-5 samples indicated that $Fe_2O_3$ have a uniform distribution on the surface of ZSM-5 zeolite and their morphologies are regular and spherical in shape.

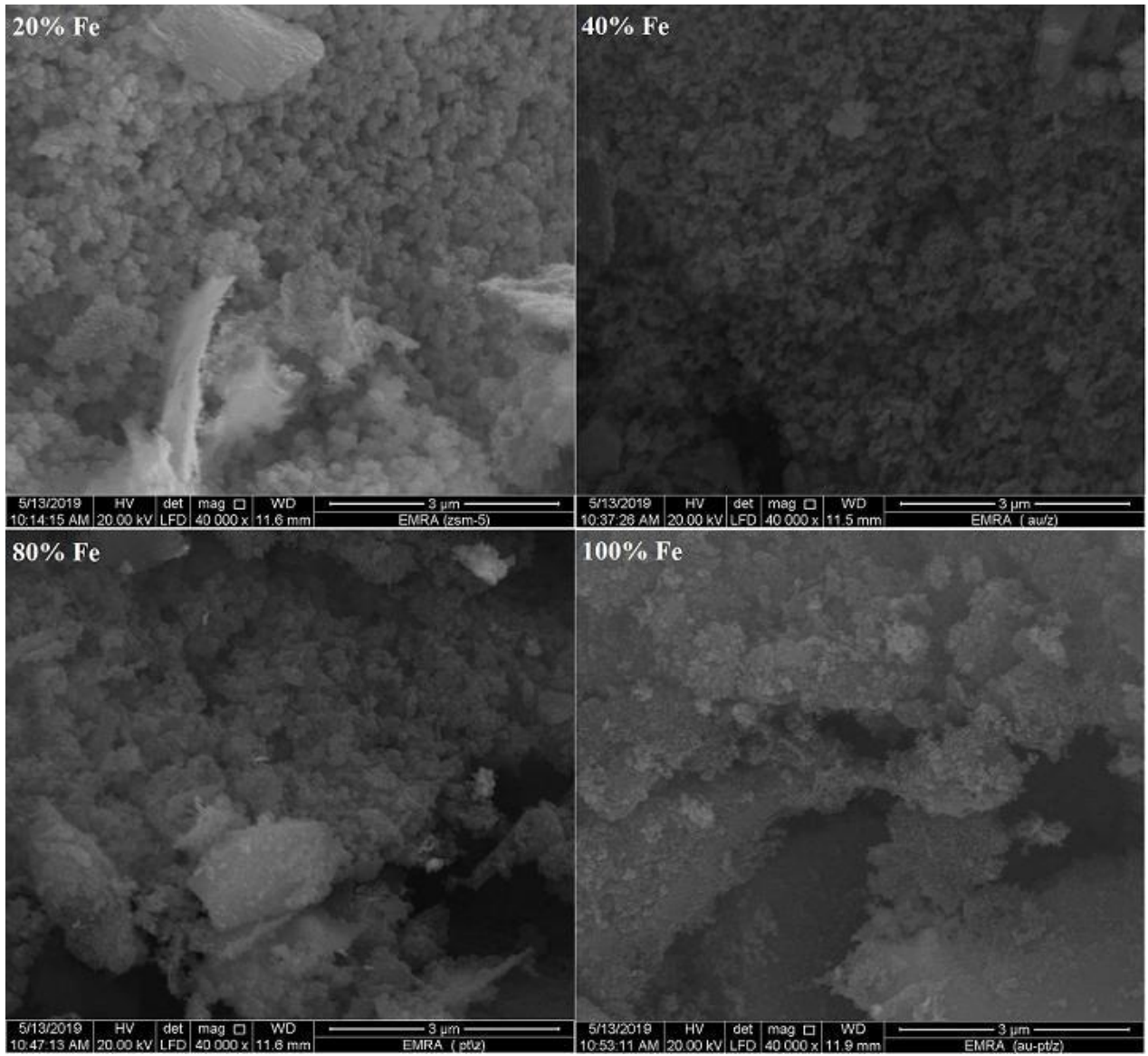

Fig. 2 SEM scans of the Fe-ZSM-5 samples

Fig. 3 presents the measured FTIR spectra of the Fe-ZSM-5 samples. Due to Fe incorporation, the spectra of all samples revealed minor changes in the typical ZSM-5 structure. The 1221 and 1083 cm$^{-1}$ absorption bands were ascribed to asymmetric stretching vibrations of TO$_4$; moreover, the bending, symmetric stretching, and double ring vibrations appeared at 792, 546 and 453 cm$^{-1}$, respectively. Additionally, increasing the iron content shifted the absorption bands at 1221(1232) and 1083 (1102) cm$^{-1}$ towards higher energies, confirming the existence of iron in the zeolite framework [13,38].

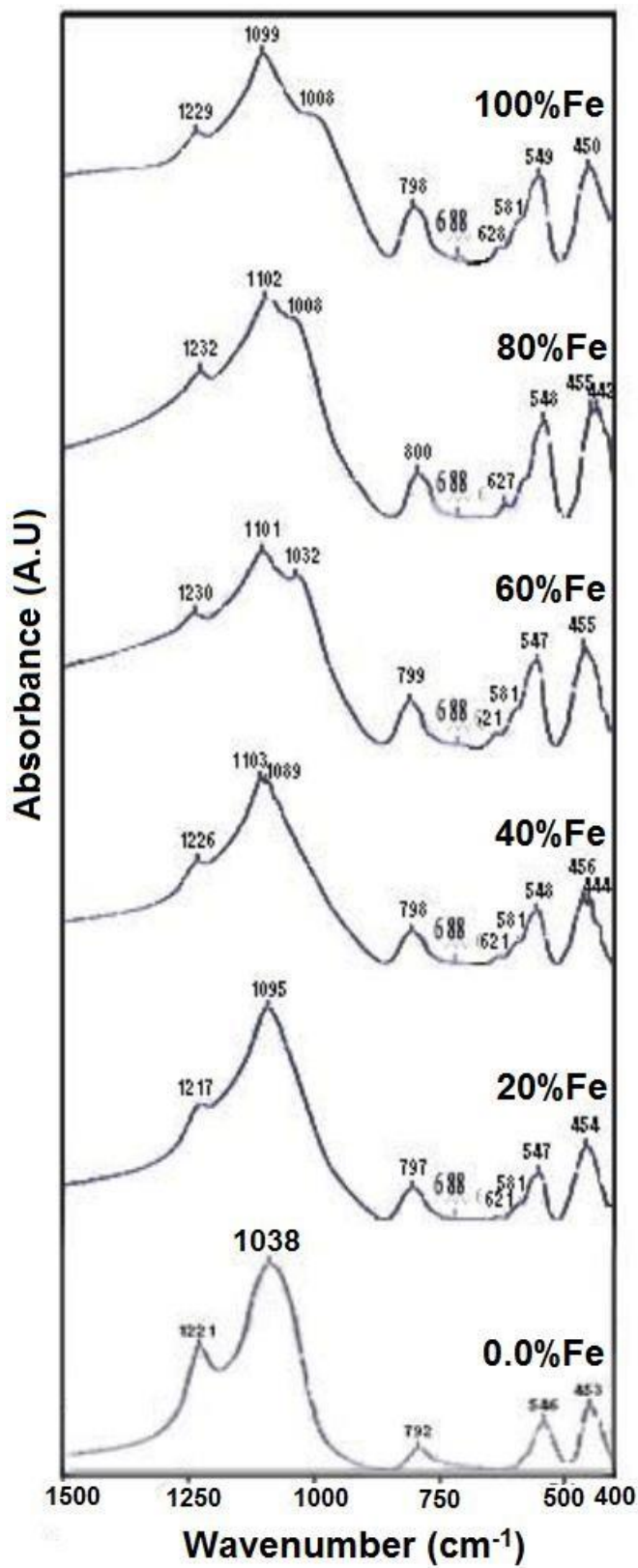

Fig. 3 Collected FTIR spectrum analysis of the Fe-ZSM-5 samples

For the parent ZSM-5, the Si-O-T internal asymmetric stretching vibration observed at 1083 cm$^{-1}$ split into 1031 cm$^{-1}$ and 1103 cm$^{-1}$ bands owing to the residual Al$^{3+}$ located near Fe$^{3+}$ at higher iron contents. As Fe–O has less ionic character (54) than Al–O (64), the former has a high covalent character and high electronegativity (1.64 vs 1.47). However, the absorption band located at 1083 cm$^{-1}$ moved to 1008 cm$^{-1}$ with increasing Fe substitution (100% Fe) because the Al-O bond is shorter than the Fe-O bond, which indicates variation in the relaxation effect and higher intensity of the Fe–O connections [13]. The Si–O-Fe symmetric stretching band located at 627 cm$^{-1}$ in the zeolite samples containing iron could be due to Fe–O vibrations, specifically to Si–O–Fe connections. In addition, at 100% Fe substitution, the intensity of the 548 cm$^{-1}$ band decreased, indicating a decrease in the crystallinity of this sample [38].

The electronic properties of Fe-ZSM-5 structures were determined from the UVVis spectral analysis shown in Fig. 4. The free Fe zeolite showed two bands at 238 and 359 nm attributed to $T_1$ and $T_2$ charge transfer processes. These transitions correspond to charge transfer between the aluminium and oxygen atoms found in zeolites. Aluminium atoms at specific locations may undergo internal redox reactions with surrounding oxygen atoms, as described in [39]:

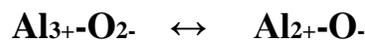

$$Al^{3+}\text{-}O^{2-} \leftrightarrow Al^{2+}\text{-}O^{-}$$

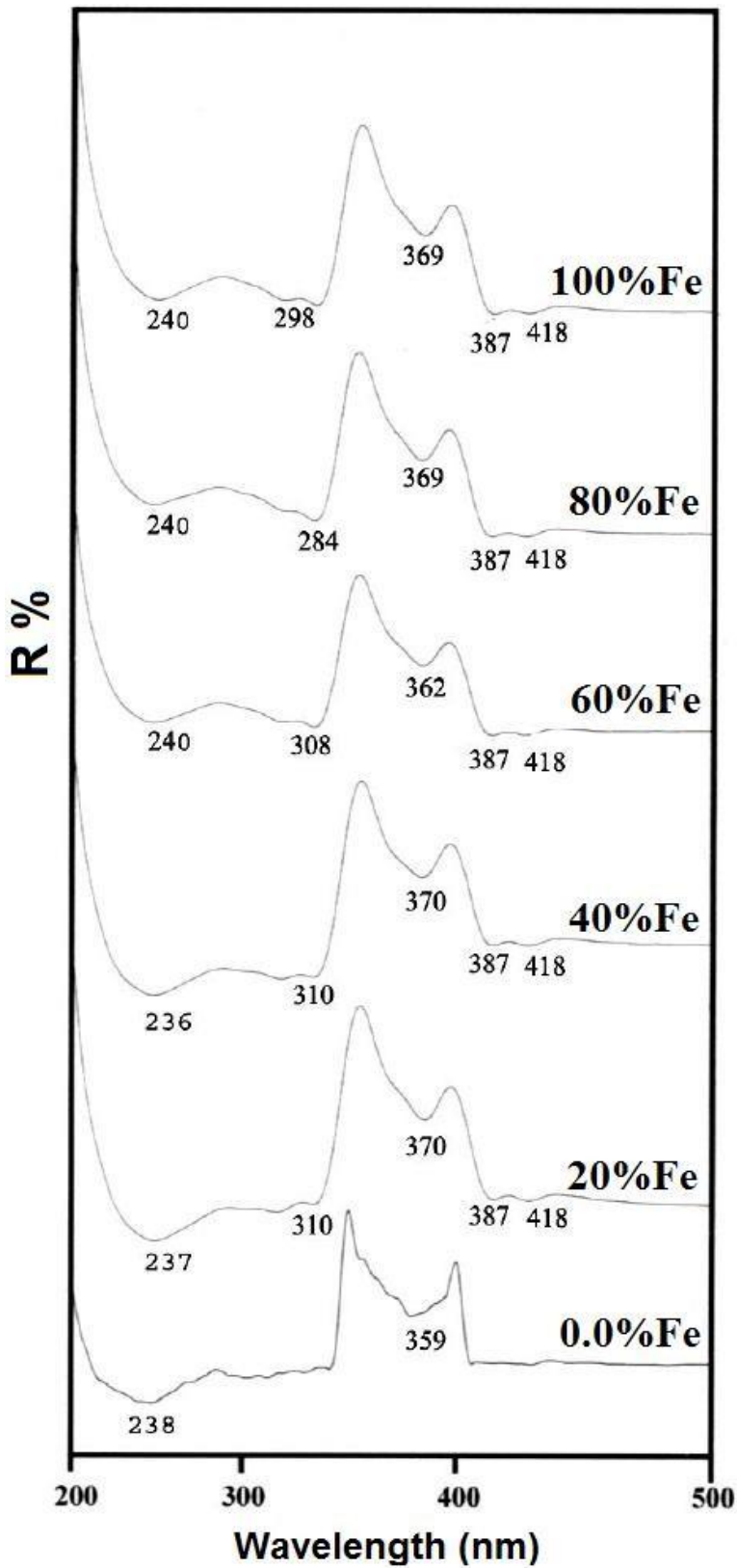

**Fig. 4** Diffused reflectance spectra of Fe-ZSM-5 samples

Zeolite samples containing iron exhibited absorption bands at 237, 310, 370, and 418 nm. The band at 237 nm is attributed to the dπ-pπ charge transfer transitions of $Fe^{3+}\leftarrow$ O in the Fe-O-Si zeolite network [39,40].

The bands in the range of 370-450 nm were attributed to the *d-d* transition of $Fe^{3+}$ ions, which shifted to shorter wavelengths with increasing iron concentrations. This finding indicates to the presence of $Fe^{3+}$ species having different coordination states other than tetrahedral $Fe^{3+}$ in $Fe_2O_3$ crystals [38]. The DSC data collected for the Fe-ZSM-5 samples are displayed in Fig. 5 and can be divided into three regions:

(i) The first region is related to desorption of water from the channels and the surfaces.

(ii) The second region corresponds to occluded template molecules present in the channels or neutralizing $SiO^-$ groups.

(iii) The temperature range of 480-550 °C illustrates the iron metal substitution by charge-balancing.

The temperature trend for desorption of $TPA^+$ with increasing substitution of Al ions by Fe decreases from 489 to 465 °C. These phenomena are related to the thermal decomposition of $TPA^+$ ions, which are easily removed with increasing Fe ion concentrations. The absence of any peaks at temperatures over 600°C in the DSC thermograms of the samples in Fig. 4 indicates the structural stability of these samples [1].

Table 2 indicates the surface characteristics of the iron zeolite samples. Notably, (i) the $S_{BET}$ and $S_t$ of the different adsorbents are similar, confirming the correct selection of the t-curve used in the analysis and indicating the absence of ultra-microporous substances; (ii) $S_{BET}$ and $V_p$ were increased by 28% and 20%, respectively, for zeolites containing 40%Fe; (iii) the average pore diameter of all zeolite samples was nearly constant; (iv) The external surface area values for the samples were ≤10% of the $S_{BET}$, revealing that the samples had a mesoporous nature, but the parent sample gives a value of 17%, reflecting higher microporosity; (v) for the sample containing 80% Fe, the total pore volume decreased compared with those of the other compositions, confirming the probable formation of iron oxide or iron silicate species [13].

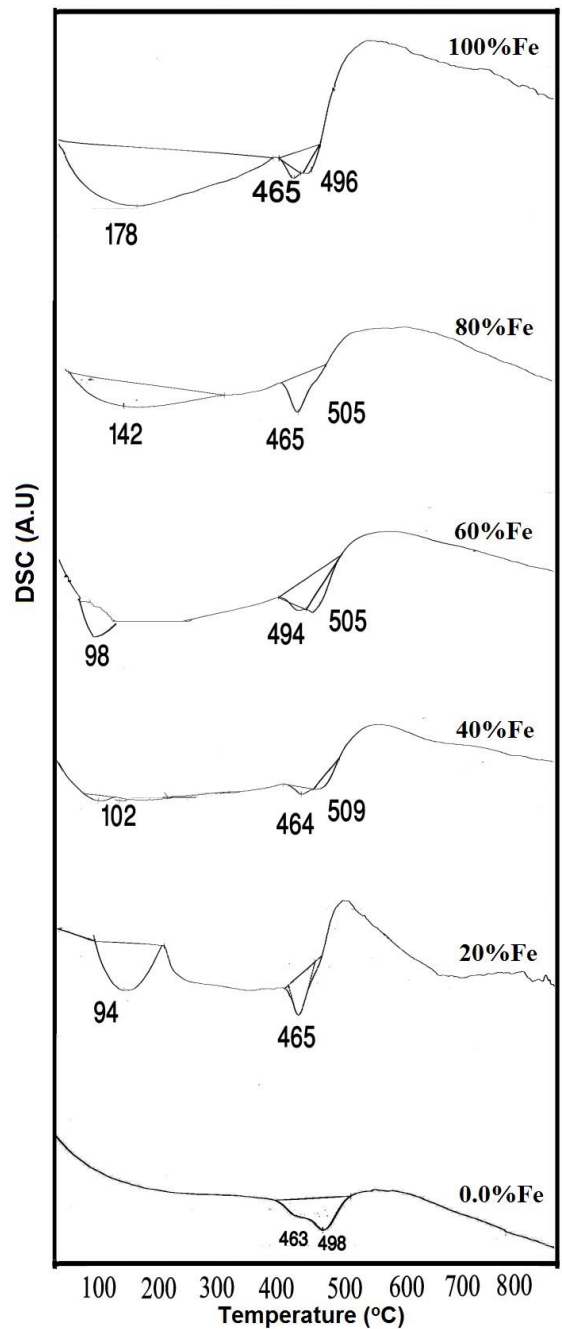

**Fig. 5** DSC curves of Fe-ZSM-5 samples

**Table 2** Surface characteristics of different Fe-ZSM-5 samples

| Samples | $S_{BET}$ (m²/g) | $S_t$ (m²/g) | $S_\mu$ (m²/g) | $S_{ext}$ (m²/g) | $S_{wid}$ (m²/g) | $\bar{r}$ (Å) | $V_{p\,total}$ (cm³/g) | $V_{p\,\mu}$ (cm³/g) | $V_{pwid}$ (cm³/g) | Micro. % |
|---|---|---|---|---|---|---|---|---|---|---|
| 0.0%Fe | 660 | 658 | 593 | 34 | 67 | 21.0 | 0.554 | 0.498 | 0.056 | 90 |
| 20%Fe | 737 | 753 | 662 | 80 | 75 | 23.3 | 0.688 | 0.617 | 0.0702 | 90 |
| 40%Fe | 846 | 840 | 788 | 75 | 56 | 20.0 | 0.696 | 0.651 | 0.0444 | 94 |

| 60%Fe | 721 | 730 | 672 | 69 | 49 | 21.3 | 0.614 | 0.572 | 0.042 | 93 |
| 80%Fe | 650 | 641 | 541 | 73 | 109 | 23.4 | 0.609 | 0.507 | 0.102 | 83 |
| 100%Fe | 668 | 660 | 593 | 65 | 75 | 26.2 | 0.698 | 0.620 | 0.0784 | 89 |

Table 3 gives the mean zone of inhibition values of ZSM-5 samples with various amounts of Fe (20% to 100%), demonstrating the excellent antibacterial activity of these samples against the tested bacteria; the inhibition value is different for each strain except for *A. fumigatus* (fungi) and *P. aeruginosa*. These results indicate that ZSM-5 effectively exhibits antibacterial activity, as do the samples with Fe (20% and 100%). Additionally, Fe-ZSM-5 nanoparticles are advantageous over the Ag nanoparticles due to the inefficient form of adsorbed Ag as silver oxide deposited on the zeolite surface (Ag NPs have poor bactericidal effect) [24].

Table 3. Mean zone of inhibition in mm ± standard deviation beyond the well diameter (6 mm) produced on a range of environmental and clinically pathogenic microorganisms using 10 mg/ml of tested sample

| Sample | ZSM-5 | 20%Fe/Z | 60%Fe/Z | 100%Fe/Z | St. |
|---|---|---|---|---|---|
| **Fungi** | | | | | **Amphotericin B** |
| *Aspergillus fumigatus* | NA | NA | NA | NA | 25.80 ± 0.10 |
| *Candida albicans* | NA | NA | 11.60 ± 0.66 | 13.40 ± 0.55 | 27.50 ± 0.10 |
| **Gram-positive bacteria** | | | | | **Ampicillin** |
| *Streptococcus pneumonia)* | NA | 12.50 ± 0.39 | 18.10 ± 0.63 | 42.40 ± 0.55 | 25.80 ± 0.20 |
| *Bacillis subtilis* | NA | 22.20 ± 0.57 | 20.90 ± 0.48 | 46.30 ± 0.67 | 37.30 ± 0.30 |
| **Gram-negative bacteria** | | | | | **Gentamicin** |
| *Pseudomonas aeruginosa* | NA | NA | NA | NA | 19.20 ± 0.10 |
| *Escherichia coli* | NA | 17.50 ± 0.27 | 22.70 ± 0.68 | 35.30 ± 0.46 | 21.90 ± 0.30 |

The mechanism of this phenomenon can be interpreted as follows. Singularized macromolecular ions accumulate at the counter (ground) electrode (Fig. 6). ZSM-5 and the samples with 20% and 100% Fe have excellent efficacy against *B. subtilis*, *S. pneumonia*, *E. coli* and *C. albicans*. However, no bactericidal ability against *A. fumigatus* or *P. aeruginosa* is observed in the test, indicating that *A. fumigatus* and *P. aeruginosa* may be tolerant bacteria and the resistance of A. fumigatus and P. aeruginosa also to most efficient Fe-ZSM-5 zeolite system as the concentration of Fe species insufficient, decreasing the bactericidal effect of nanoparticles [24]. In comparison, unmodified ZSM-5 did not produce any inhibitory effect on any kind of bacteria or fungi because it does not release metal ion species.

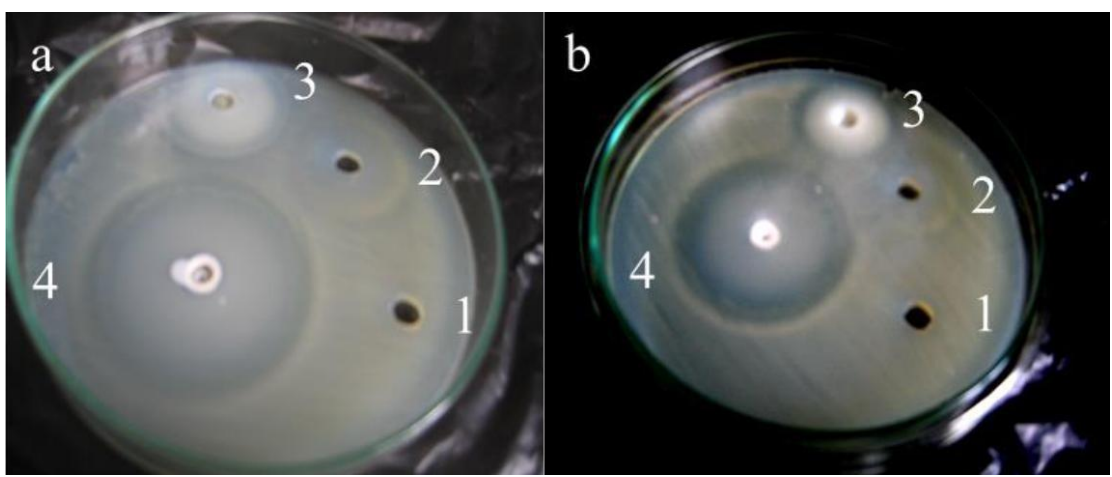

Fig. 6 Antibacterial tests of ZSM-5 (1) in comparison to 20% Fe/Z (2), 605Fe/Z (3) 100% Fe/Z (4) against *Bacillis subtilis* (gram positive) (a) and *Escherichia coli* (gram negative) (b) bacteria as examples

4. Conclusions

A hydrothermal route was applied to silica extracted from RHS to prepare the Fe-ZSM-5 zeolite. The chemical and physical changes on the Fe-ZSM-5 surfaces were characterized via XRD, FTIR, diffuse reflectance, DSC, and pore structure analysis. XRD analysis revealed the typical ZSM-5 structure with new diffraction lines of the iron silicate phase. A new Si–O–Fe band at 656 cm$^{-1}$ appeared for all Fe-ZSM-5 samples, as confirmed by the FTIR spectra. XRD and UV-Vis spectral analysis indicated the presence of tetrahedral $Fe^{3+}$. The antibacterial activity of such coatings towards different kinds of bacteria and fungi, such as *S. pneumonia*, *B. subtilis*, *E. coli*, *P. aeruginosa*, *A. fumigatus*, and *C. albicans*, for the investigated ZSM-5 and Fe (20% and 100%) samples showed selective antibacterial actions. Most antibacterial activity was observed against *B. subtilis* and *E. coli*. ZSM-5 alone or with Fe modification has no effect on the inhibition of such bacteria and fungi. Moreover, no bactericidal ability against *A. fumigatus* and *P. aeruginosa* was observed in the test, indicating that *A. fumigatus* and *P. aeruginosa* may be tolerant bacteria.

On behalf of all authors, the corresponding author states that there is no conflict of interest.

***Declaration of Interest Statement**

**Declaration of interests**

☒ The authors declare that they have no known competing financial interests or personal relationships that could have appeared to influence the work reported in this paper.

☐ The authors declare the following financial interests/personal relationships which may be considered as potential competing interests: